# Inverse-designed lithium niobate nanophotonics

Chengfei Shang,[1†] Jingwei Yang,[1†] Alec M. Hammond,[2] Zhaoxi Chen,[1] Mo Chen,[3] Zin Lin,[4] Steven G. Johnson,[3*] and Cheng Wang[1*]

[1]*Department of Electrical Engineering & State Key Laboratory of Terahertz and Millimeter Waves, City University of Hong Kong, Kowloon, Hong Kong, China*
[2]*School of Electrical and Computer Engineering, Georgia Institute of Technology, Atlanta, GA 30308, USA*
[3]*Department of Mathematics, Massachusetts Institute of Technology, Cambridge, MA 02139, USA*
[4]*Department of Electrical and Computer Engineering, Virginia Polytechnic Institute and State University, Arlington, VA 22203, USA*

[†]*These authors contributed equally to this article*

[*]*stevenj@math.mit.edu*
[*]*cwang257@cityu.edu.hk*

*Abstract:* Lithium niobate-on-insulator (LNOI) is an emerging photonic platform that exhibits favorable material properties (such as low optical loss, strong nonlinearities, and stability) and enables large-scale integration with stronger optical confinement, showing promise for future optical networks, quantum processors, and nonlinear optical systems. However, while photonics engineering has entered the era of automated "inverse design" via optimization in recent years, the design of LNOI integrated photonic devices still mostly relies on intuitive models and inefficient parameter sweeps, limiting the accessible parameter space, performance, and functionality. Here, we develop and implement a 3D gradient-based inverse-design model tailored for topology optimization of the LNOI platform, which not only could efficiently search a large parameter space but also takes into account practical fabrication constraints, including minimum feature sizes and etched sidewall angles. We experimentally demonstrate a spatial-mode multiplexer, a waveguide crossing, and a compact waveguide bend, all with low insertion losses, tiny footprints, and excellent agreement between simulation and experimental results. The devices, together with the design methodology, represent a crucial step towards the variety of advanced device functionalities needed in future LNOI photonics, and could provide compact and cost-effective solutions for future optical links, quantum technologies and nonlinear optics.

# Introduction

The lithium niobate-on-insulator (LNOI) platform has seen rapid development[1] in recent years and shows great potential in future advanced photonics systems, owing to the excellent optical properties of lithium niobate (LN), including large nonlinear susceptibilities, a wide optical transparency window, and great stability, as well as strong optical confinement that allows compact and scalable integrated photonic devices and circuits to be built on wafer scales. A wide range of on-chip nanophotonic devices, including high-speed electro-optic modulators[2][3][4], high-Q micro resonators[4][6], broadband frequency comb generators[7][8][9], efficient frequency convertors[10][10][12][13], entangled photon pair generators[14], spectrometer[15], optical isolator[16], ultrafast all-optical switches[17] and frequency shifters[18] have been demonstrated, making LN-based photonic integrated circuits a promising solution for future high-speed optical networks, quantum information processing, and nonlinear optics. However, to date the design of most LNOI devices still relies heavily on human intuition augmented by simple analytical models, and can only access limited parameter spaces by scanning a few hand-selected parameters (such as widths, gaps and radii of curvature), either manually or through inefficient parameter sweeps (whose cost scales exponentially with the number of parameters).

Inverse-design methods[19], on the other hand, have recently attracted considerable attention in nanophotonics, which automatically search for the optimal topological structure of a pre-specified objective in a certain design region, exploring a large parameter space through high-dimensional optimization algorithms. Inverse design can reveal highly non-intuitive device designs with extremely compact sizes (several micrometers in diameter) and unprecedented functionalities. Many inverse-designed functional devices have been demonstrated in recent years, such as mode multiplexers and convertors[20][21], wavelength multiplexers[22][23], meta-surfaces[24][25], nonlinear wavelength convertors[25][26][27], dispersion-engineered microresonators[28][29], together with system applications such as massively parallel optical transmitters[30], particle accelerators[31], and chip-based light detection and ranging (LiDAR) systems[32][33].

Thus far, most inverse-designed photonic devices have been demonstrated in silicon (Si) and other CMOS-compatible platforms such as silicon nitride[24] because of the mature fabrication technologies available for those materials. Compared with Si photonics, the challenges in the design and implementation of inverse-designed LNOI devices mainly arise from practical fabrication constraints. Due to the difficulty in dry etching LN and electro-optic overlap considerations, typical waveguides in LNOI feature a rib structure with an unetched slab underneath, further reducing the already smaller effective index contrast compared with that of Si photonics. Moreover, dry-etched LNOI waveguides usually exhibit a substantial sidewall angle, which needs to be taken into consideration during the optimization process to achieve accurate modeling, and which also limits the minimally achievable feature sizes. More recently, inverse-design algorithms that take into account specific geometric constraints[34] and sidewall angles[35] have led to designs and demonstrations compatible with standard foundry services[36] and in more exotic material platforms such as diamond[37] and silicon carbide[29]. However, the realization of inverse-designed LNOI devices is still missing and could substantially benefit the development of compact LNOI photonic integrated circuits, especially when dealing with complex design problems with multiple objective figures of merit.

Here, we overcome these fabrication and design challenges and demonstrate a series of compact high-performance inverse-designed LNOI photonic devices based on the open-source Meep package[38]. The design algorithm takes into full consideration of the fabrication constraints in the LNOI platform, including the rib structures, minimum feature sizes and etched sidewalls. We design and fabricate a spatial-mode multiplexer that separates the fundamental transverse-electric ($TE_0$) mode and second-order TE ($TE_1$) mode, with an insertion loss ~ 1.5 dB and a crosstalk < -15.8 dB, a waveguide crossing with a low loss of 0.48 dB and a crosstalk < -36 dB, and a compact waveguide bend that turns the propagation of light by 90° with a radius of curvature of 6 μm and a loss of 0.41 dB. The devices show excellent agreement between theoretical and experimental performances, as well as broad operation bandwidths from 1500 nm to 1600 nm wavelength.

## Results

Figure 1a shows an overview of our inverse design strategy specially tailored for the LNOI platform and a representative iteration curve. The inverse design relies on a hybrid time/frequency-domain topology optimization algorithm[39] with adjoint sensitivity analysis and multiple constraint functions. Starting with a homogeneous design region (typically initialized to "gray"—halfway between air and LNOI) and a given objective function, e.g., the transmission coefficients of a particular waveguide mode, the optimization solver efficiently calculates the objective performance and the gradients in each iteration step by two FDTD (finite-difference time domain) simulations regardless of the number of pixelated design parameters. Compared with optimization problems in SOI, the slanted sidewalls in the LNOI platform require special optimization algorithm design, whereas the existence of an unetched slab substantially increases the difficulty in achieving good optimization performances due to a lowered effective index contrast. In this paper we use a 400 nm z-cut LN device layer with a 250 nm etch depth and a sidewall angle of 45 degree for all designs, similar to other devices previously demonstrated in our group[4][40].

The optimization strategy shown in Fig. 1a contains three main steps as outlined in Ref. 39. In the first step, permittivity in the design region is allowed to vary continuously, while a convolution and projection function is introduced to smoothen the design region and improve the dynamic range. The projection strength $\beta$ is increased gradually during the iterations to avoid a dramatic change in the design parameters, such that the optimizer focuses on improving the figure of merit while gradually pushing the design towards a more binarized permittivity distribution. In the second step, we incorporate a geometric constraint[34] to eliminate features smaller than the indicated minimum length scales, during which the design parameters are further binarized. Finally, we introduce sidewall features by using linearly shifted threshold values for the projection function at different vertical slices[35], so that the pattern is "eroded" linearly with height above the slab surface (Fig. 1b). In each step, the optimizer runs for several tens of iterations before convergence, the precise number depending on the specific problem. Importantly, all simulations are performed in 3D, taking fully into account the practical rib/slab thicknesses, crystal anisotropy, and sidewall angles, which are crucial to realize experimentally achievable designs. More mathematical details regarding the inverse design algorithms can be found in Methods and previous references[34][35][41]. Figure 1c-e shows schematic views of the designed structures and their corresponding functions of our mode multiplexer, waveguide crossing and compact waveguide bend, respectively.

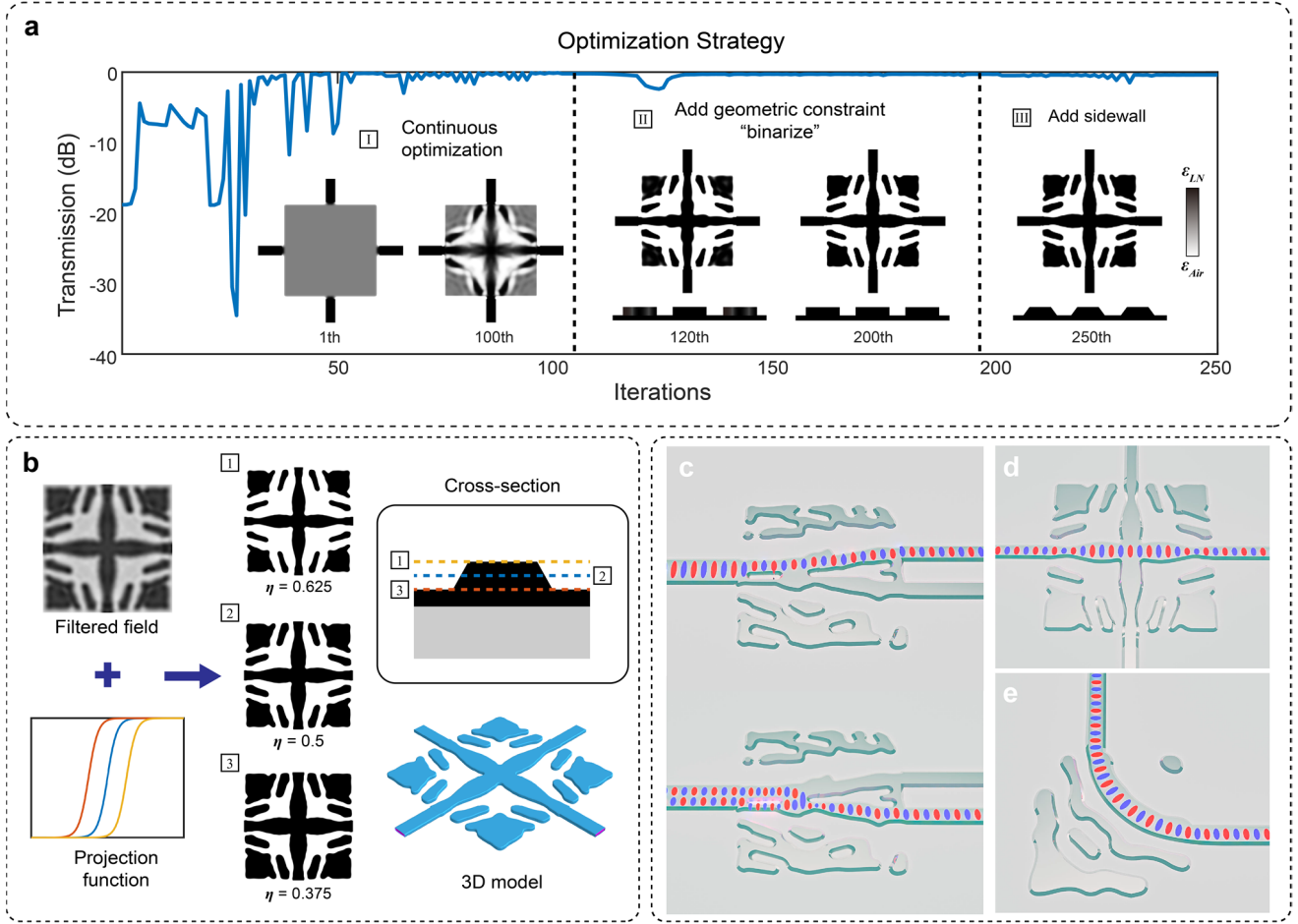

**Fig. 1.** Overview of the optimization strategy and device schematic. **a,** General optimization strategy and a typical iteration curve of a waveguide crossing optimization problem consisting of three main steps. In the first step, the permittivity in each pixel is allowed to vary continuously; second, we add geometric constraints to binarize the structure and achieve a minimum length scale; finally, slanted sidewall is introduced while the solver reoptimizes the design to "compensate" for the sidewall-induced performance degradation. The cross-section views in step 2 and step 3 are drawn along the diagonal of the design region. **b,** Method of adding sidewall features. We introduce linearly shifted threshold values $\eta$ for the hyperbolic tangent projection function at different heights (left), which leads to eroded ($\eta = 0.625$), normal ($\eta = 0.5$) and dilated ($\eta = 0.375$) structures at the top, center and bottom slices of the rib structure, respectively (right). We design and fabricate **c,** a $TE_0/TE_1$ mode multiplexer, **d,** a waveguide crossing, and **e,** a compact waveguide bend to verify the design strategy, all of which show good performance and excellent fabrication compatibility.

## Spatial Mode Multiplexer

First, we consider an LNOI spatial mode multiplexer, an important component for mode-division multiplexing (MDM) technology in future high-volume data transmission systems[42]. In our design, the fundamental $TE_0$ mode in a 2 μm wide input waveguide will be coupled into the upper output arm, whereas the $TE_1$ mode will be converted into $TE_0$ mode, and output from the bottom arm (Fig. 2a). The two output waveguides are both 1 μm wide and are separated by a 2 μm gap. Figure 2a shows the final inverse-designed pattern together with simulated field evolution ($E_y$) for $TE_0$ and $TE_1$ input, whereas Figure 2b shows the scanning electron microscope (SEM) image of the fabricated mode multiplexer. The footprint of the final design is $12 \times 12$ μm$^2$, orders of magnitude smaller compared with mode multiplexers based on traditional asymmetrical directional couplers (ADCs)[43] or cascaded Mach–Zehnder interferometers (MZI)[44]. The minimum feature size is set to be 0.2 μm (at the middle slice of the rib) to balance between the potential degrees of freedom in design and the fabrication constraints. Our simulation results show that, for both $TE_0$ and $TE_1$ input scenarios, a single mode multiplexer features an average insertion loss less than 0.9 dB and crosstalk less than -17 dB over a 100-nm wavelength window (1500 nm-1600 nm). In our actual experiments, we fabricate two multiplexers back to back connected by a 50-μm-long, 2-μm-wide multimode waveguide, such that all input/output signals are in $TE_0$ mode and are easier to be precisely calibrated. The simulation result (Fig. 2c) for such a cascaded mode multiplexer pair shows average insertion losses at the desired outputs ($S_{31}$ and $S_{42}$) of less than 2 dB (twice the loss of a single device), and crosstalk ($S_{41}$ and $S_{32}$) of less than -16.5 dB. In the experimental test (Fig. 2d), we measure average insertion losses of less than 3 dB and crosstalk less than -15.8 dB for the mode-multiplexer pair within a wavelength range between 1520 nm and 1600 nm, consistent with the simulation results. The measured insertion loss values are further corroborated by comparing the losses of two and four cascaded mode-multiplexers (Methods). From the results we estimate that a single mode multiplexer features an insertion loss ~ 1.5 dB and an upper bond crosstalk value of -15.8 dB within the 80-nm wavelength window (see Methods for a detailed analysis). The remaining differences between simulation and experimental values could result from deviations in the actually fabricated device parameters and fabrication imperfections especially on the small features.

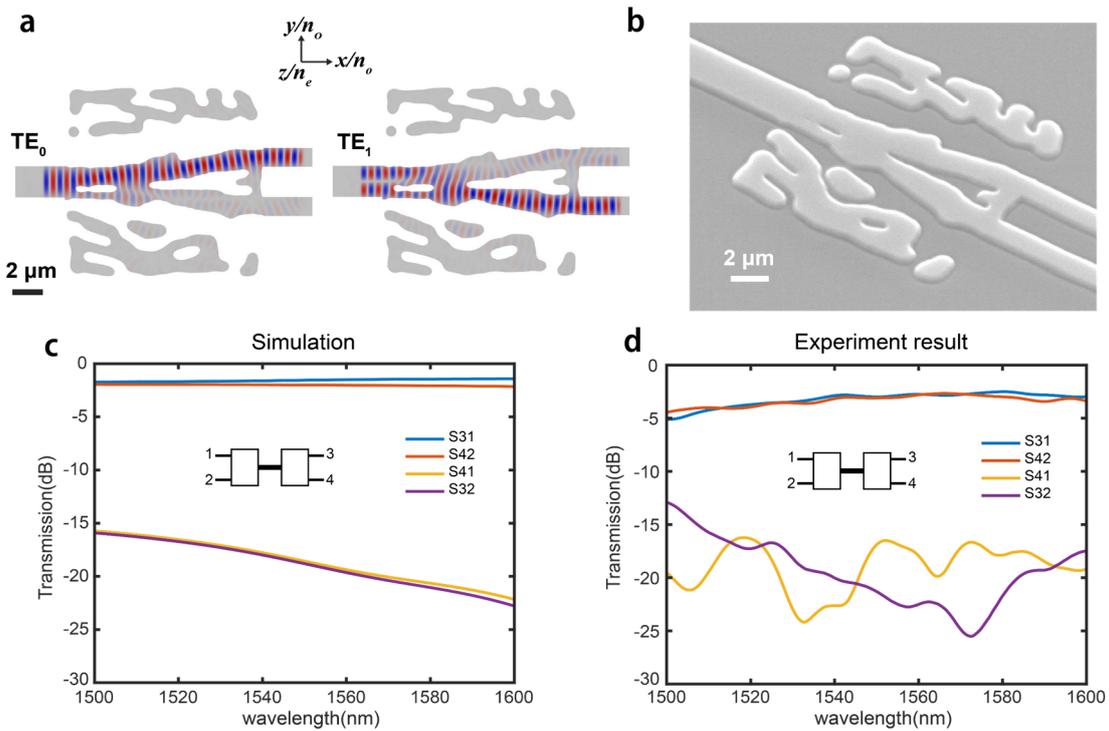

**Fig. 2.** Inverse-designed spatial mode multiplexer. **a,** The optimized design pattern and simulated field ($E_y$) evolution of the mode multiplexer with a 12 × 12 μm² footprint. The gray and white areas correspond to LN rib and slab regions, respectively. **b,** SEM image of the fabricated mode multiplexer. **c,** Simulated transmission coefficients of a back-to-back mode multiplexer pair between the four input/output ports as shown in the inset. **d,** Experimentally measured transmission coefficients of a fabricated mode multiplexer pair, showing broadband low-loss and low-crosstalk operation consistent with simulation results.

## Waveguide crossing

A waveguide crossing[45][46][47][48][49][50] is an essential component for signal routing in large-scale and high-density photonic integrated circuits. Traditional waveguide crossing designs typically rely on heuristic shaped taper or multimode interferometer (MMI) structures combined with exhaustive parameter sweeps[48][49]. Here we design a compact waveguide crossing using our inverse-design algorithm without the need for an initial guess. The design region (Fig. 3a) again has a footprint of 12 × 12 μm$^2$, and is set to have mirror symmetry along both horizontal and vertical directions, since the *x*- and *y*-crystal orientations are isotropic in our z-cut LNOI wafer. The objective function is designed to maximize the power of fundamental TE$_0$ mode. Here the minimum feature size is set to be 0.5 μm as this optimization problem is easier to converge. Figure 3b shows the SEM image of the fabricated waveguide crossing. The simulated average insertion loss over a 100-nm wavelength band (Fig. 3c) is 0.22 dB, with low crosstalk of less than -40 dB. In the experiment, we measure the insertion loss by cascading five and twenty crossing structures and comparing with a single waveguide crossing (Fig. 3d), showing a low fitted average insertion loss of 0.48 dB per crossing over the tested wavelength range (1500-1600 nm). The crosstalk measured from the two vertical output waveguides (port 3 and port 4 as indicated in the inset of Fig. 4d) is -36 dB and -39 dB, respectively, consistent with the simulation results.

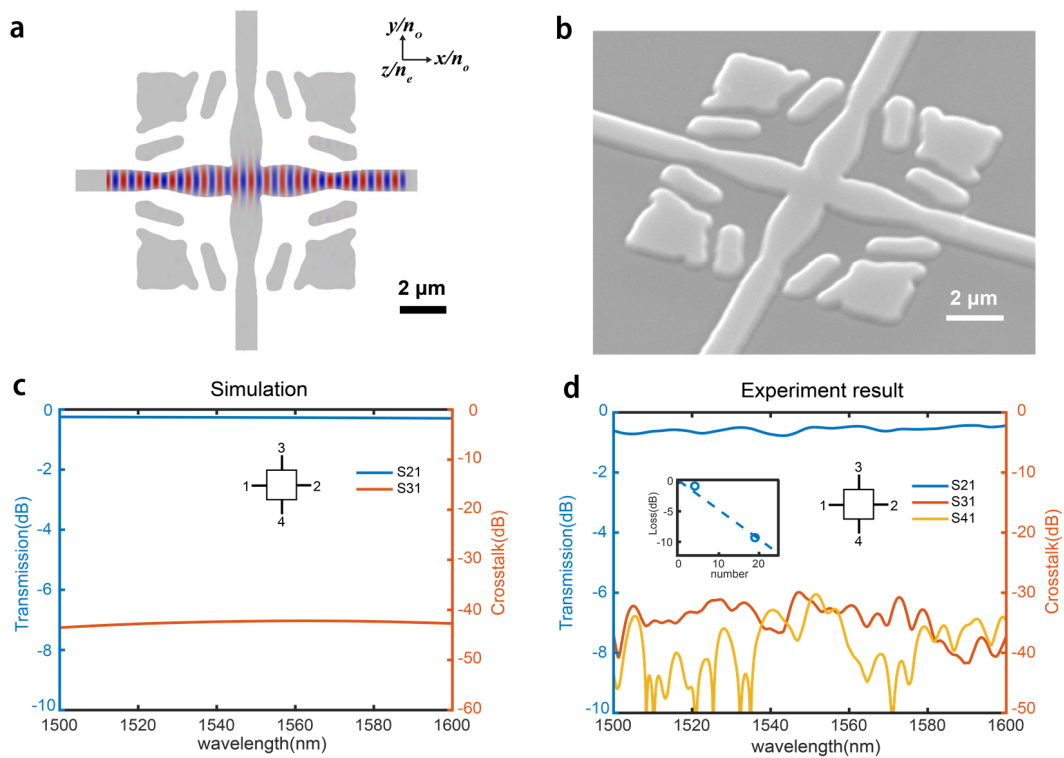

**Fig. 3.** Inverse-designed LNOI waveguide crossing. **a,** The optimized design pattern and simulated field ($E_y$) evolution of the waveguide crossing with a 12 × 12 μm$^2$ footprint. The gray and white areas correspond to LN rib and slab regions, respectively. **b,** SEM image of the fabricated crossing device. **c,** Simulated optical transmission (S$_{21}$) and crosstalk (S$_{31}$) of the designed waveguide crossing. **d,** Experimentally measured device performance, showing a low insertion loss and crosstalk from 1500 nm to 1600 nm. Inset shows cut-back loss measurement results of 5 and 20 cascaded crossing structures.

## Compact waveguide bend

Finally, we demonstrate a compact bending waveguide that rotates the propagation direction of the fundamental TE$_0$ mode by 90° within a tight bending radius of 6 μm. Due to the existence of unetched slab, waveguide bends in LNOI platforms usually require radii of at least 30 μm to limit the radiation loss. A simple circular 90° bend with a radius of 6 μm in our current platform will lead to a high loss of 3.7 dB according to our numerical simulation. Here the design region (Fig. 4a) is 10 × 10 μm$^2$ which is nearly an order of magnitude smaller than a traditional 30-μm bend, both widths of input and output waveguides are 1 μm, and the minimum length scale is 0.4 μm. Figure 4b shows the SEM image of the fabricated waveguide bend. Physically, this optimized bend design could be roughly interpreted as a sharp bend augmented with a Bragg-mirror-like structure to suppress radiation loss, qualitatively similar to bends designed by topology optimization in other material platforms[51], but determining the precise details requires the power of inverse design. The simulated average transmission loss (Fig. 4c, blue) in the desired band (1500 nm~1600 nm) is 0.29 dB, which is equivalent to a simple bend with a radius of 30 μm according to our simulation. In the experimental test, we cascaded two, four, and six bends and compare the measured total insertion losses with a reference waveguide (Fig. 4c, red and inset). The average measured insertion loss for a single bend is 0.41 dB over the 100-nm wavelength range. The measured loss is slightly higher than expected, possibly due to an under-etched rib height (~ 230 nm) resulting in more power leakage through the thicker slab. Nevertheless, the measured loss is still more than 8 times lower than that of a simple circular bend with the same radius.

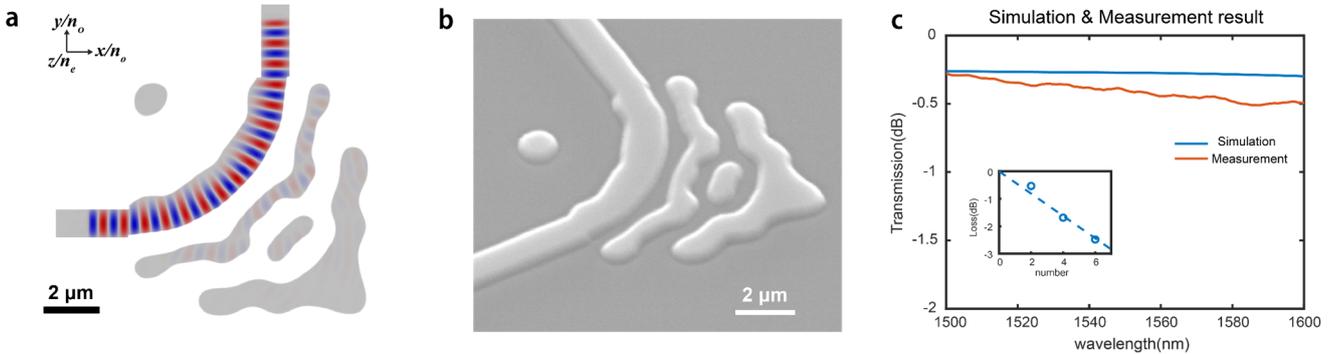

**Fig. 4.** Inverse-designed LNOI waveguide bend. **a**, The optimized design pattern and simulated field ($H_z$) evolution of the bending waveguide with a 6-μm radius. The gray and white areas correspond to LN rib and slab regions, respectively. **b**, SEM image of the fabricated waveguide bend. **c**, Simulated and experimentally measured optical transmission of the waveguide bend, showing a low average insertion loss from 1500 nm to 1600 nm. Inset shows cut-back loss measurement results of two, four and six cascaded bends.

## Conclusions and outlook

In summary, we have successfully developed an inverse-design algorithm compatible with the practical fabrication constraints in the LNOI platform, including non-vertical sidewalls, relatively large minimum length scales and an unetched-slab structure. Based on this algorithm, we demonstrate, for the first time, a series of inverse-designed LNOI devices with small footprints, low losses, low crosstalk, broad bandwidths, and good consistency with theoretical prediction. Further relaxing the minimum length constraints could lead to designs compatible with stepper photolithography processes[52] with much better scalability and cost-effectiveness. The same design methodology could be readily applied to achieve LNOI devices with more advanced functions, especially those make use of the electro-optic and/or nonlinear properties of LN, such as fast-tunable switches[17], dispersion-engineered comb generators[8], and efficient nonlinear wavelength convertors[10], and be extended to other material platforms that share similar fabrication constraints, such as yttrium orthovanadate[53] and titanium dioxide[54]. The devices, together with the design methods, could become important building blocks for future LNOI functional photonic circuits with applications in on-chip optical links, quantum technologies and nonlinear optics.

## Methods

### Device design

The inverse design algorithm in this work uses a gradient topology optimization based on objective functions calculated using a FDTD solver (Meep). Adjoint method is used to calculate the gradients in each optimization iteration, which obtains all gradient information in each pixel of the design region by two separate forward and backward FDTD calculations. A general inverse design problem with design variables $\boldsymbol{\rho}$ (Eq. (1-1)) is described as follows:

$$min_\rho \ t \quad (1\text{-}1)$$

$$\text{s.t.} \ \nabla \times \frac{1}{\mu_0} \times \boldsymbol{E} - \omega_n^2 \mu_0 \varepsilon_0 \varepsilon_r(\widetilde{\widetilde{\boldsymbol{\rho}}})\boldsymbol{E} = -i\omega_n \boldsymbol{J} \quad (1\text{-}2)$$

$$\boldsymbol{\varepsilon}_r(\widetilde{\widetilde{\boldsymbol{\rho}}}) = \boldsymbol{\varepsilon}_{air} + \widetilde{\widetilde{\boldsymbol{\rho}}}(\boldsymbol{\varepsilon}_{air} - \boldsymbol{\varepsilon}_{LN}) \quad (1\text{-}3)$$

$$0 \le \widetilde{\widetilde{\boldsymbol{\rho}}} \le 1 \quad (1\text{-}4)$$

$$f(\omega_n) - t \le 0 \quad (1\text{-}5)$$

$$g_k \le G_k t \quad k \in \{LS, LW\} \quad (1\text{-}6)$$

where Eq. (1-2) is the physical rule of the optimization (i.e. Maxwell equation), Eqs. (1-3&4) convert the final design field $\widetilde{\widetilde{\boldsymbol{\rho}}}$, which is projected from the design variables $\boldsymbol{\rho}$ (to be discussed later) and is bounded between 0 and 1, into the actual permittivity in the material grids via a linear interpolation ($\boldsymbol{\varepsilon}_{air} \le \boldsymbol{\varepsilon}_r(\widetilde{\widetilde{\boldsymbol{\rho}}}) \le \boldsymbol{\varepsilon}_{LN}$). Since all optimization problems in this work deal with minimax-type objectives (e.g. maximizing the minimum transmission at 10 wavelength points), a dummy parameter $t$ is introduced to transform the minimax problems to be differentiable. $t$ is linked with the multiple objective functions to be optimized via a set of constraint functions (1-5), where $f(\omega_n)$ is the error from an ideal performance, which translates the minimax problem to a single differentiable optimization problem of minimizing $t$. Eq. (1-6) is the geometric constraint function $g_k$, which is dependent on the minimum linewidth and spacing (LS & LW) in the design region. $G_k$ indicates the constraint boundary, which is dynamically tightened or loosened according to the real-time value of the dummy parameter, i.e. the geometric constraint is set tighter when the performance is closer to the desired value. Details of the gradient-based adjoint optimization method can be found in Ref. [34].

The design variables $\boldsymbol{\rho}$ are translated into the final field $\widetilde{\widetilde{\boldsymbol{\rho}}}$ via an extra mapping step, which smoothens the design region to remove single-pixel features, binarizes the pattern, and add possible symmetry constraints. In the optimization, we calculate the gradient $\frac{\partial FOM}{\partial \rho}$ from $\frac{\partial FOM}{\partial \widetilde{\widetilde{\rho}}}$ via the back-propagation method after adjoint simulation. A symmetric design could be realized by performing a mirror or rotation operation to the design field and averaging with the original field:

$$\boldsymbol{\rho}_{symm} = average(\boldsymbol{\rho} + \boldsymbol{\rho}_{flip} + \boldsymbol{\rho}_{rotate} + \cdots) \quad (1\text{-}7)$$

where $\boldsymbol{\rho}_{symm} = \boldsymbol{\rho}$ if no symmetry is present. The mapping function then utilizes a linear-convolution filter to achieve a smoothened field $\widetilde{\boldsymbol{\rho}}$:

$$\widetilde{\boldsymbol{\rho}} = w(\boldsymbol{x}) * \boldsymbol{\rho}_{symm} \quad (1\text{-}8)$$

The final field $\widetilde{\widetilde{\boldsymbol{\rho}}}$ is finally obtained by a nonlinear thresholding projection function[35]:

$$\widetilde{\widetilde{\rho}} = \frac{tanh(\beta \cdot \eta) + tanh(\beta \cdot (\widetilde{\rho} - \eta))}{tanh(\beta \cdot \eta) + tanh(\beta \cdot (1 - \eta))} \tag{1-9}$$

where the $\beta$ parameter determines the thresholding sharpness and $\eta$ is the threshold value. The slanted sidewall feature is added to our design problem by introducing a height-dependent projection function where the projection threshold $\eta(z)$ varies in z direction as:

$$\eta(z) = \eta_1 + \left(\frac{h-z}{h}\right)(\eta_2 - \eta_1) \tag{1-10}$$

where $\eta_2 < \eta_1 \in [0,1]$ and $h$ denotes the height from the surface of the slab, which induces a linear "erosion" effect from bottom to top of the design region when the design field is projected to the final field, as Fig. 1b shows.

### Performance estimation of a single mode multiplexer

The performance of a single spatial mode multiplexer (demultiplexer) could be estimated from the measured S-parameters of one and multiple back-to-back multiplexer pairs in our experiment. According to the structural symmetry and system reciprocity, the insertion losses of a single multiplexer are simply half of the measured losses of the multiplexer pair ($S_{31}$ for $TE_0$ and $S_{42}$ for $TE_1$). We have also tested the insertion losses of two pairs of multiplexers with measured total insertion losses of 6.18 dB for $S_{31}$ and 6.21 dB for $S_{42}$, corroborating the estimated 1.5 dB insertion loss for a single multiplexer. The crosstalk of a multiplexer pair is mainly attributed from two paths ($S_{41}$ as an example): (i) unwanted crosstalk in the first multiplexer into $TE_1$ mode ($c_1$, which corresponds to field coefficient) that couples out from the bar port of the second multiplexer ($t_2$); and (ii) light passes through the first multiplexer into $TE_0$ mode as desired ($t_1$) but goes into the cross port of the second multiplexer ($c_2$). As a result, $S_{41} = |c_1 t_2 + e^{i\varphi} t_1 c_2|^2$, where $\varphi$ is the phase difference induced by the waveguide between the two multiplexers. A similar analysis would show that $S_{32}$ should be represented by the same expression. Since our devices feature high overall transmission levels, i.e. $t_1 \approx t_2 \approx 1$, the measured crosstalk values $S_{32}$ and $S_{41}$ are approximately equal to $|c_1|^2 + |c_2|^2$ on average since the two paths experience different phases at different wavelengths, which is the sum of and provides an upper bound for the two actual crosstalk values (for port 1 and port 2 input) of a single mode multiplexer. If we assume the two crosstalk values are the same ($|c_1| = |c_2|$), the actual crosstalk of a single multiplexer should be 3 dB lower than that of the measured multiplexer pair, i.e. -18.8 dB.

### Device fabrication & characterization

Devices are fabricated from commercially available 400 nm thick z-cut LN thin films (NANOLN). Inverse design patterns are first defined in hydrogen silsesquioxane (HSQ) using an electron-beam lithography (EBL, Crestec 9510C, 50 keV) system, and then transferred into the LN layer using optimized argon plasma-based reactive ion etching (RIE, Plasma-Therm Version 320). The LN etch depth is 250 nm, leaving a 150-nm-thick slab. The final devices are cleaved for end-fire coupling. For optical characterizations, light from a continuous-wave tunable telecom laser (Santec TSL-710) is sent to the devices under test using a lensed fiber after a polarization controller to ensure TE polarization. The output optical signal is collected using a second lensed fiber and sent to a 125-MHz photodetector (New Focus 1811) to obtain the optical transmission spectra.

# Acknowledgement

We thank Dr. Wenzhao Sun for valuable discussions on the manuscript. This work is supported in part by National Natural Science Foundation of China (61922092), Research Grants Council, University Grants Committee (CityU 11204820, CityU 11212721, N_CityU113/20), Croucher Foundation (9509005), and by a grant from the Simons Foundation.